\documentclass[11pt]{article}

\usepackage[utf8]{inputenc}
\usepackage{authblk}      
\usepackage[most]{tcolorbox}
\usepackage{float}
\usepackage{graphicx}
\usepackage{caption}
\usepackage{natbib}
\usepackage{algpseudocode}
\usepackage{amsmath}
\usepackage{amssymb}
\usepackage{url}
\DeclareMathOperator*{\argmax}{arg\,max}

\title{Community Detection with the Canonical Ensemble}
\author[1]{Rudy Arthur\thanks{Corresponding author: \texttt{R.Arthur@exeter.ac.uk}}}
\affil[1]{Department of Computer Science, University of Exeter, Exeter EX4 4PY, UK}

\date{\today}

\begin{document}

\maketitle

\begin{abstract}
 \noindent 
 Network community detection is usually considered as an unsupervised learning problem. Given a network, the aim is to partition it using some general purpose algorithm. In this paper we instead treat community detection as a hypothesis testing problem. Given a network, we examine the evidence for specific community structure in the observed network compared to a null model. To do this we define an appropriate test statistic, analogous to a z-score, and several null models derived from maximising entropy under different constraints in the canonical ensemble. We demonstrate the application of this method on real and synthetic data and contrast our method to Bayesian approaches based on the stochastic block model. We demonstrate that this method gives definitive answers to concrete questions, which can be more useful to analysts than the output of a generic algorithm. 
\end{abstract}

\section{Introduction}\label{sec:introduction}
 
Community detection involves partitioning the nodes of a network into groups. Generally communities are defined such that nodes are densely connected within a community and sparsely connected between them \citep{radicchi2004defining}. Communities are the most important of the various `meso-scale' structures which can be used to summarise network data, though core-periphery \citep{kojaku2017finding}, disassortative \citep{chen2014anti} and other patterns are also studied \citep{peixoto2014hierarchical, liu2023nonassortative, arthur2025detectability}.

Community detection in networks shares many similarities with clustering vector data, with numerous methods for transforming one problem to the other \citep{white2005spectral, von2007tutorial}. Community detection and clustering also share a number of `philosophical' issues regarding the interpretation of the output of a community detection/clustering algorithm \citep{kleinberg2002impossibility}. In this paper we approach community detection from the point of view of \cite{von2012clustering} that \emph{`clustering should not be treated as an application-independent mathematical problem, but should always be studied in the context of its end-use'}.

This means, rather than presuming there is a single correct partition of a network which some universal algorithm can discover, community detection should provide a way to probe a network with questions motivated by the origin and intended application of the data.  We do this in the framework of statistical hypothesis testing, which is well established in community detection \citep{lancichinetti2010statistical, lancichinetti2011finding, kojaku2018generalised}. Hypothesis testing compares the value of some test statistic to the distribution of that statistic under the null. If the observed value is very unlikely to be generated by the null (quantified by the p-value) then we can reject the null and claim that the statistic is `significant' i.e. the network has some structure not explained by the null.

Newman's modularity function \citep{newman2006modularity} or some variant of it, is often used as the test statistic \citep{massen2006thermodynamics,zhang2017hypothesis,kojaku2018generalised,yanchenko2024generalized}. However modularity uses a null model which is often invalid, it only detects assortative community structures and it is not properly normalised. We will discuss these issues in the next section and the first contribution of this work is a test statistic that can be used instead of modularity, correcting these issues. We can use this statistic to detect communities or any other `block structure' so patterns like assortative/disassortative/core-periphery and others can be tested. It also allows the possibility of unassigned nodes, something common in vector clustering approaches \citep{ester1996density,ankerst1999optics,campello2013density} but rare in network clustering algorithms. 

The second contribution of this paper is a thorough consideration of null models appropriate for community detection. We derive null models by maximising entropy under constraints. This gives various functions which may be used to in different scenarios but, to be clear, this paper is \textbf{not} simply proposing new functions to maximise or a way to `fix' modularity. Rather we claim that a rigorous approach to community detection requires more than simply maximising a quality function. To this end we provide a practical scheme to ask and answer questions about the meso-scale structure of networks. These questions should be motivated by the origin and application of the data. The aim is for a more thoughtful and rigorous approach to community detection in applications, rather than current practice which often involves throwing the network into the author's preferred black-box algorithm then performing a \emph{post hoc} analysis of the returned partition. 

In Section \ref{sec:genzmod} we define our test statistic. In Section \ref{sec:null} we derive our null models through entropy maximisation with constraints. Section \ref{sec:algo} describes the community detection procedure and implementation details. In Section \ref{sec:artificial} we study different artificial examples, showing that this method agrees with theoretical expectations for community detectability. We compare our approach to the method of Stochastic Block Model (SBM) inference in Section \ref{sec:SBM} and show how these different approaches can lead to different answers about the same network. The correct approach depends on what kind of data analysis that is being undertaken, which we discuss in our concluding Section \ref{sec:conclusion}.

\section{Generalised Z-Modularity}\label{sec:genzmod}

Modularity is usually defined as
\begin{equation}\label{eqn:newmanmod}
    Q_{Newman}(G,c) = \frac{1}{2E}\sum_{i}^N \sum_{j}^{N} \left(A_{ij}(G) - \frac{k_ik_j}{2E}\right) \delta( c(i), c(j) )
\end{equation}
where $N$ is the number of nodes in the network, $A_{ij}(G)$ is the adjacency matrix of the network $G$, $k_i = \sum_{j} A_{ij}$ is the degree of node $i$, $2E = \sum_i k_i$ is twice the number of edges and $c(i)$ is the labelling function that returns the `community' of node $i$ and defines a partition of the network.

The term $\frac{k_i k_j}{2E}$ is the approximate probability for two nodes to be connected in the configuration model of $G$, that is, if the degrees of the nodes are fixed and the connections randomised, the probability of an edge between $i$ and $j$ is $P_{ij} \simeq \frac{k_i k_j}{2E}$.  This is valid only in the sparse limit, where no node has degree greater than $\sqrt{N}$. \cite{garlaschelli2017ensemble} note the correct expression, valid always, is well-known \citep{park2004statistical} and ought to be used instead (see Section \ref{sec:null}). Other null models such as a random, Erd\H{o}s--R\'enyi, graph where $P_{ij} = \frac{2E}{N(N-1)}$ or more complicated examples \citep{young2007random, garlaschelli2004fitness} could also be substituted. Allowing for an arbitrary null model gives,
\begin{equation}
    Q(G,P,c) = \frac{1}{2E}\sum_{ij} \left(A_{ij} - P_{ij}\right) \delta( c(i), c(j) )
\end{equation}
where $P_{ij}$ is the probability of connecting $i$ and $j$ in the null.

Following \citep{arthur2023discovering}, we can use a straightforward generalization of modularity to detect structures other than assortative communities. Defining the block variables
\begin{align}
    S_{ab}(G, c) &= \sum_{i \in a, j \in b} A_{ij}(G) \\
    T_{ab}(c) &= \sum_{i \in a, j \in b} P_{ij} \\
    Q_{ab} &= S_{ab} - T_{ab}
\end{align}
we have
\begin{equation*}
    Q(G,P,c) = \frac{1}{2E} \sum_{a}^K Q_{aa}
\end{equation*}
where $K$ is the number of groups. We introduce the $K \times K$ `block matrix' $B_{ab}$, which specifies the meso-scale structure we are searching for and define 
\begin{equation}
    Q_{block}(G,P,B,c) = \frac{1}{2E} \sum_{a}^K \sum_b^K B_{ab} Q_{ab}
\end{equation}
For example, if $B$ has $1$s on the diagonal and $0$s elsewhere the above reduces to $Q_{Newman}$. As shown in \citep{arthur2023discovering, arthur2025detectability, arthur2025exploring} other patterns in the $B$ matrix can be used to detect other meso-scale structures.

In hypothesis testing the difference between observed and expected outcomes is usually measured relative to the variance of the null. This missing normalisation of standard modularity is responsible for the tendency, discussed by \citep{peixoto2023descriptive}, of partitions maximising $Q_{Newman}$ to favour equal sized groups. Similarly to \cite{miyauchi2016z} we normalise $Q$ by the standard deviation of the null. The variance of the expected edge count between blocks $a$ and $b$, assuming independent links, is
\begin{equation}
    Var_{ab}(P_{ij}) = \sum_{i \in a, j \in b} P_{ij} (1 - P_{ij}) = T_{ab} - U_{ab}
\end{equation}
with
\begin{equation}
    U_{ab} = \sum_{i \in a, j \in b} P_{ij}^2
\end{equation}
We then define
\begin{equation}
    Z_{ab} = \frac{S_{ab} - T_{ab}}{\sqrt{Var_{ab}}} = \frac{S_{ab} - T_{ab}}{\sqrt{T_{ab} - U_{ab}}}
\end{equation}
and finally
\begin{equation}\label{eqn:canmod}
    Z(G,P,B,c) = \frac{1}{ C } \sum_{a}^C \sum_b^C B_{ab} Z_{ab}(G,P,c)
\end{equation}

The difference between $Z$ and \cite{miyauchi2016z}'s Z-modularity (apart from the possibility of non-assortative structures and different null models) is that here we normalise each group separately. If the groups are large and the labels randomly assigned each $Z_{ab} \sim \mathcal{N}(0,1)$ and summing over $C^2$ groups gives $Z \sim \mathcal{N}(0,1)$. Thus each group contributes equally to the total variance, regardless of group size. This method of combining independent z-scores is called Stouffer's method in the literature on meta-analysis \citep{owen2009karl}.

\subsection{Unassigned Nodes}\label{sec:unassigned}

A pattern to detect two communities would be
$$
B = \begin{pmatrix}
    1 & -1\\
    -1 & 1
\end{pmatrix}
$$
Rewarding an excess of edges between nodes with the same label and a deficit of edges between nodes with different labels. If we add a row and columns of $0$s
$$
B = \begin{pmatrix}
    1 & -1 & 0\\
    -1 & 1 & 0\\
    0 &0 & 0
\end{pmatrix}
$$
we now have 3 groups. The quality function $Z$ with this block matrix rewards an excess of edges within and a deficit between the first two groups. Nodes in the last group makes no contribution to the score. Therefore, nodes which can't be placed in either of the first two groups without decreasing the score can be placed in the third. These nodes do not interact with the community structure and we call this the `unassigned' group.

We could use any values for $B_{ab}$ to design a test statistic that searches for structures of interest. In this paper, for simplicity, we just use values $B_{ab} = \pm 1$ to specify structure and an extra row and column of $0$s when we want to allow unassigned nodes. See \citep{arthur2025detectability} for more discussion of possible patterns for the $B$ matrix.

\section{Null Models}\label{sec:null}

The test statistic, $Z$, depends on the null model, $P$. Different choices affect what communities we can identify in a network, or indeed, if we can identify any at all. A null model assigns a probability $P(G)$ to a network $G$. This is often discussed in statistical mechanical language where we consider ensembles of networks \citep{park2004statistical,bianconi2009entropy}. If networks in the ensemble are subject to constraints, for example that they should have some fixed number of edges or a specified degree sequence, we determine $P(G)$ by maximising the entropy subject to those constraints.

It is important to distinguish constraints that are satisfied exactly from constraints satisfied in an ensemble average, corresponding to the microcanonical and canonical ensembles respectively. As argued by \cite{garlaschelli2017ensemble}, we typically want to use the canonical ensemble, allowing for the fact that edges in our observed network may be imprecisely measured. When the number of constraints is extensive, which is the case in the configuration model and most of the models we consider, then the microcanonical and canonical ensembles are not equivalent \citep{squartini2015breaking}.

We first write down the general recipe for computing the edge probability in the canonical ensemble and describe specialisations for specific nulls below. Assuming independent Bernoulli edges with connection probability $P_{ij}$, the probability of generating a (binary, undirected, simple) network with adjacency matrix $A$ is
\begin{equation}
    P(A) = \prod_{i < j} P_{ij}^{A_{ij}} (1-P_{ij})^{1 - A_{ij}}
\end{equation}
The entropy is 
\begin{align}
    S &= -\sum_{A} P(A) \log P(A) \\
    &= -\sum_A P(A) \sum_{i<j} A_{ij} \log P_{ij} + (1-A_{ij})\log(1-P_{ij})\\ \nonumber
    &= -\sum_{i < j} P_{ij} \log P_{ij} + (1-P_{ij})\log(1-P_{ij})\\
\end{align}
where $\sum_A$ indicates a sum over the ensemble of all possible networks and we used
\begin{equation}
    P_{ij} = \sum_A P(A) A_{ij}
\end{equation} 
The entropy can be maximised with constraints via the method of Lagrange multipliers, 
\begin{align}
    {\cal S} &= S + \sum_\mu \lambda_\mu \left(  C_\mu(P)  - \bar{C}_\mu \right)
\end{align}
where the constraints $C_\mu(P)$ are some functions that must have an expected values equal to $\bar{C}_\mu$. Maximising ${\cal S}$ gives,
\begin{align}
    \log \frac{P_{ij}}{ 1-P_{ij}}   = \sum_\mu \lambda_\mu \frac{ \partial C_\mu}{\partial P_{ij}}
\end{align}
Using
\begin{align*}
    \sigma(x) = \frac{1}{1-e^{-x}},  \qquad \sigma^{-1}(y) = \log \frac{y}{1-y}
\end{align*}
gives
\begin{align}\label{eqn:edgeprob}
    P_{ij}  = \sigma( \sum_\mu \lambda_\mu \frac{ \partial C_\mu}{\partial P_{ij}} )
\end{align}
The parameters $\lambda_\mu$ are fixed by substituting $P$ into the constraint equations and solving. 

If the constraints are linear in $P$ (as they will be in all of our examples) then
\begin{align}\label{eqn:linearconstraints}
    C_\mu &= \sum_{ij} P_{ij} c_{\mu ij} \rightarrow \partial_{ij} C_\mu = c_{\mu ij} + c_{\mu ji}
\end{align}
writing $\partial_{ij} C_\mu \equiv \frac{\partial C_\mu}{\partial P_{ij}}$ for brevity. It can be seen (by differentiating) that maximising
\begin{equation}\label{eqn:maxent}
    {\cal L} = \sum_\mu \lambda_\mu\bar{C}_\mu  - \sum_{i<j} \log\left(1 + \exp\left( \sum_\mu \lambda_\mu \partial_{ij} C_\mu \right) \right)
\end{equation}
is equivalent to solving the constraint equations.

This form is almost equivalent to maximising the likelihood of a Bernoulli random edge model. The log likelihood of such a model is,
\begin{equation}\label{eqn:BernoulliL}
    \log L(\lambda, \alpha | A) = \sum_{i<j} A_{ij} \log P_{ij} + (1-A_{ij})\log(1- P_{ij})
\end{equation}
Substituting $P_{ij}$ from equation \ref{eqn:edgeprob} and using
\begin{align*}
    \log \sigma(x) = x - \log(1 + e^x), \qquad \log (1 - \sigma(x)) = -\log(1 + e^x)
\end{align*}
gives
\begin{equation}\label{eqn:BernoulliL2}
    \log L(\lambda, \alpha | A) = \sum_{i<j} A_{ij}  \sum_\mu \lambda_\mu \partial_{ij} C_\mu   - \sum_{i<j} \log\left(1 + \exp\left( \sum_\mu \lambda_\mu \partial_{ij} C_\mu \right) \right)
\end{equation}
which is equivalent to ${\cal L}$ when the constraints are in the form of Equation \ref{eqn:linearconstraints} and
\begin{equation}
        \bar{C}_\mu = \sum_{ij} A_{ij} c_{\mu ij} 
\end{equation}
In which case the constraint equations are
\begin{equation}
        \sum_{ij} A_{ij} c_{\mu ij} =  \sum_{ij} P_{ij} c_{\mu ij} 
\end{equation}

\subsection{Erd\H{o}s--R\'enyi Model}

In this model the number of edges is constrained to equal the number in the observed network
\begin{align}
   C(P) &= \sum_{ij} P_{ij} = 2E
\end{align}
Substituting into the above formulas, where in this case we can solve the constraint by hand, gives
\begin{align}
   P_{ij} =  \sigma( \lambda) = p = \frac{2L}{N(N-1)}
\end{align}
Testing against this null we are asking the question \emph{is there community/meso-scale structure in the network beyond what would be expected in a random network with the same number of nodes and edges?}

\subsection{Configuration Model}

Here the degrees are constrained to equal the observed degrees, on average. This gives $N$ equations
\begin{align}
 \sum_j P_{ij} = \sum_j A_{ij}
\end{align}
Substituting into the above formulas gives
\begin{align}
   P_{ij} = \sigma( \lambda_i + \lambda_j )
\end{align}
And the multipliers $\lambda_i$ are found by maximising
\begin{equation}\label{eqn:configlike}
  {\cal L} = \sum_i k_i \lambda_i - \sum_{i < j} \log(1 + e^{\lambda_i + \lambda_j})    
\end{equation}
We will refer to this as the canonical configuration model. Testing against this null we are asking \emph{is there community/meso-scale structure in the network beyond what would be expected from the node degrees?}

\subsection{Random Dot Product Graph}
The configuration model is a relatively weak null model. Many networks likely to be of interest in practice will have more structure than their degree distribution alone implies. A stricter null can be constructed using a Random Dot Product Graph (RDPG) \citep{young2007random}. We consider a low-rank approximation for the adjacency matrix 
\begin{equation}
    A \simeq V V^T
\end{equation}
where the $N$ rows of $V$ are $d$-dimensional vectors, $v_m$. Standard results in linear algebra suggest that $v_m = \sqrt{\mu_m} e_m$, where $e_m$ are the largest normalised eigenvectors of $A$ with eigenvalues $\mu_m$, will provide a good approximation. 

Requiring that the null and the observed network have the same low-rank approximation gives $Nd$ equations
\begin{align}\label{eqn:lowrankconstraint}
 \sum_j P_{ij} v_{jm}  &= \sum_j A_{ij} v_{jm}
\end{align}
We also require that the edge count is the same
\begin{align}
    \sum_{ij} P_{ij} = \sum_{ij} A_{ij}
\end{align}
Substituting into the maximum entropy formulation we get,
\begin{align}\label{eqn:rdpgnull}
   P_{ij} =  \sigma \left( \lambda + \sum_m \lambda_{mi} v_{jm} + \sum_m \lambda_{mj} v_{im} \right) 
\end{align}
The multipliers can be found by maximising
\begin{equation}\label{eqn:rdpglik}
  {\cal L} =  \sum_{i<j}A_{ij}(\lambda + v_{jm} \lambda_{mi} + v_{im} \lambda_{mj}) - \sum_{i < j}\log(1 + e^{\lambda + \lambda_{mi} v_{jm} + \lambda_{mj}v_{im}} ) 
\end{equation}
 In this equation, the vectors $v_{*m}$ are assumed to be known and fixed to equal the largest $d$ eigenvectors of $A$. We maximise Equation \ref{eqn:rdpglik} to find the parameters $\lambda_{mi}$. This model is derived in alternative way in \citep{o2020maximum}. Note that the canonical configuration model is a special case of the RDPG where we use a rank one approximation $v = \mathbf{1}$, a vector with all entries equal to 1.

Testing against this null we are asking the question \emph{is there community structure in the network beyond what is present in the low-rank approximation?} It is important to note that failure to achieve significance with this null does not mean the absence of community structure, it means that whatever structure is there is already accounted for by the low-rank approximation.

\subsection{Gravity Model}

Spatial networks often show many connections between spatially close nodes and few direct connections between distant nodes. One specific null for spatial data which encodes this tendency is the gravity model \citep{garlaschelli2004fitness, di2022gravity}. We start from the model of \citep{expert2011uncovering} which suggests, in the sparse limit,
\begin{equation}\label{eqn:gravity}
    P_{ij} = \delta w_i w_j f(d_{ij})
\end{equation}
where $w_i$ is some known importance value associated with each node e.g. population, $f(d)$ is the deterrence function which depends on the physical separation between nodes $d_{ij}$ and $\delta$ is a proportionality constant.

Requiring that the degree be proportional to the importance gives
\begin{equation}\label{eqn:degreew}
    \sum_j P_{ij} = \delta w_i
\end{equation}
Requiring that the number of edges be preserved gives
\begin{equation}
    \sum_{ij} P_{ij} = \sum_{ij} A_{ij} = 2E
\end{equation}
Which fixes $\delta$ since
\begin{equation}
    \sum_{ij} P_{ij} = \delta \sum_i w_i = 2E
\end{equation}
so $\delta = 2E/W$ with $W = \sum_i w_i$.

If we ignore deterrence ($f(d_{ij}) = 1$) the degree constraint gives
\begin{equation}
    P_{ij} = \sigma(\lambda_i + \lambda_j) 
\end{equation}
Often the identification $e^{\lambda_i} = \sqrt{\delta} w_i$ is made based on the sparse limit \citep{garlaschelli2004fitness}. We could instead impose it as a constraint, by maximising 
\begin{equation}\label{eqn:gravpop}
    {\cal L} = \sum_i \delta w_i  \lambda_i - \sum_{ij} \log(1 + e^{\lambda_i +\lambda_j})
\end{equation}
as usual. This is not a maximum likelihood estimate but it is a maximum entropy one. External constraints, i.e. not only involving network data $A_{ij}$, do not appear in the likelihood, Equation \ref{eqn:BernoulliL2}, but they can fit in the maximum entropy model, Equation \ref{eqn:maxent}. In practice, using this method the values $w_i$ cannot be too far from the degrees $k_i$, otherwise it might be impossible to satisfy the constraints or find a maximum for Equation \ref{eqn:gravpop} and we will ignore node importance values in the rest of this work, substituting $w_i = k_i$ and $\delta = 1$.

We introduce distance dependence using the constraint from \cite{expert2011uncovering} that the number of edges in each distance `bin' is fixed
\begin{equation}\label{eqn:distconst}
    \sum_{ij} P_{ij} I_b(d_{ij}) = \sum_{ij} A_{ij} I_b(d_{ij})
\end{equation}
where the bin indicator function for bin $b$
\begin{equation}
    I_b(d_{ij}) = \begin{cases}
        1 \quad d_{ij} \in b\\
        0 \quad \text{otherwise}
    \end{cases}
\end{equation}
and the bins are assumed to be given. The sum of edges is constrained automatically, since every edge is in some bin. These constraints together with the degree constraints give
\begin{equation}\label{eqn:gravitynull}
    P_{ij} = \sigma\left( \lambda_i + \lambda_j + \sum_b \lambda_b I_b(d_{ij})\right) 
\end{equation}
Where the parameters are found by maximising
\begin{equation}
    {\cal L} = \sum_i \delta w_i  \lambda_i + \sum_b \sum_{i<j}  A_{ij} I_b(d_{ij}) \lambda_b - \sum_{i<j} \log(1 + \exp( \lambda_i +\lambda_j + \sum_b \lambda_b I_b(d_{ij}))
\end{equation}
Testing against this null we are asking the question \emph{is there community structure in the network beyond what is implied by the node importances and the spatial distribution of edges?} 

\section{Algorithm}\label{sec:algo}
To perform `community detection' on some network $G$ we carry out the following procedure
\begin{enumerate}
    \item Formulate a research hypothesis. Operationally, this means specifying the null model, $P$, and a target structure $B$, which also means choosing the number of groups, $K$.
    \item Using equation \ref{eqn:canmod} find
    \begin{equation}
            \hat{Z}(G,P,B) = \max_c Z(G,P,B,c)
    \end{equation}
    This is the partition of the observed network which has the largest z-score.
    \item Generate $N_P$ random networks, $G_p$, choosing to connect each of the $N(N-1)/2$ unique pairs $ij$  with probability $P_{ij}$ \citep{squartini2015unbiased}.
    \item For all $N_P$ generated graphs find
    \begin{equation}
        \hat{Z}_p(G_p,P,B) = \max_c Z(G_p,P,B,c)
    \end{equation}
    \item Estimate the right-tailed p-value by comparing $\hat{Z}$ to the $\hat{Z}_p$ values. If the p-value is less than some predefined threshold, say $\alpha = 0.01$, reject the null and declare that the network does have structure beyond the null. Otherwise we can conclude that the observed network structure is found in many networks in the null ensemble. In this sense we can say the structure is `explained' by the constraints.
\end{enumerate}

We find the Lagrange multipliers by performing the maximisation using the LGBFS algorithm \citep{liu1989limited}\footnote{\url{https://lbfgspp.statr.me}}. Each evaluation of the likelihood function and its derivatives is $O(N^2)$. Generating a network from the null is also $O(N^2)$. We find $\hat{Z} = \max_c Z(G,P,B,c)$ by label swapping with simulated annealing, as described in \citep{arthur2023discovering}. Each label swap proposal requires $O(N)$ operations so each pass through the entire network is again $O(N^2)$.

The scaling is $O(N^2)$, rather than $O(E)$, because all of the $P_{ij}$ values are potentially non-zero. In practice, many of them are very close to zero. Thus we could apply a threshold and `sparsify' the $P$ matrix, giving us an operation count closer to $O(E)$. We did not do this here since it was not necessary for the small examples studied.

Without sparsification $O(N^2)$ scaling means this method is not yet practical for very large networks. We leave speeding up the method for future work, but note that many real problems in network analysis are  small to `medium' data problems on, say, $\sim100$ to $\sim 1000$ nodes. The examples below all run reasonably quickly on modest hardware\footnote{seconds to minutes on Intel(R) Core(TM) i5-1145G7 @ 2.60GHz}, so for such problems our approach is quite feasible.

\section{Artificial Networks}\label{sec:artificial}

\subsection{Planted Partition Model}

 The planted partition model (PPM) \citep{condon2001algorithms} is a special case of the Stochastic Block Model (SBM). Given a labelling function $c$ on $N$ nodes, the SBM generates a network with edge probability $P_{ij} = p_{c_i c_j}$, where the values $p_{ab}$ control the probability of edges within or between groups. The PPM has all the diagonal probabilities equal to $p_{in}$ and all the off-diagonal probabilities equal to $p_{out}$, so $p_{ab} = p_{out} + \delta_{ab}(p_{in} - p_{out})$. 

The degree corrected stochastic block model (dc-SBM) \citep{karrer2011stochastic} allows for more realistic degree distributions than the standard version. To generate networks from a degree corrected Planted Partition Model (dc-PPM) we sample $N$ values $t_i$ and block normalise them to get parameters $\theta_i$. In practice we will sample the $t_i$ from a Pareto distribution with exponent $\gamma$. We then choose values of $\omega_{aa} = \omega_{in}$ and $\omega_{a\neq b} = \omega_{out}$ and use a Poisson model where the edge $ij$ is added with probability $1 - \exp(-\theta_i \theta_j \omega_{g_i g_j})$. To be clear, \emph{we are only using the SBM to generate test networks}, we are not trying to fit the parameters as is done in approaches to community detection based on the SBM \citep{peixoto2014hierarchical}. We apply the steps of Section \ref{sec:algo} to study the detected community structure and compare it to the planted structure.

\subsection{ER and Canonical Nulls}
\begin{figure}[H]
    \centering
    \includegraphics[width=\textwidth]{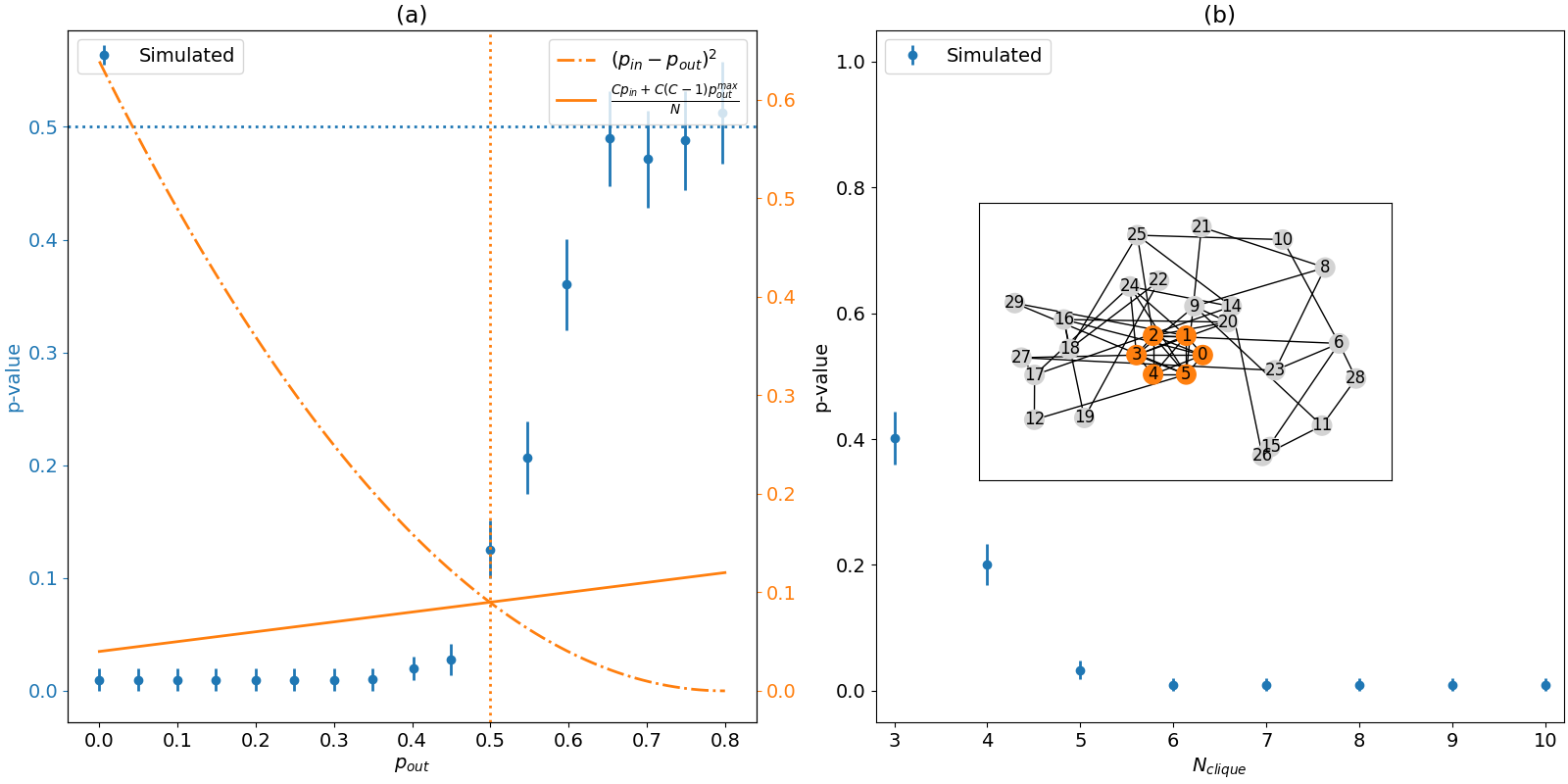}
    \caption{(a) p-values for detection of a 3 community PPM under the ER null model. The vertical dashed line shows where the Kesten-Stigum bound is broken. The horizontal dashed line shows a p-value of 0.5. (b) p-values for ER model + clique, inset figure shows the detected community for an example network with $N_{clique} = 6$. Grey nodes are unassigned. }
    \label{fig:er}
\end{figure}
First we study the standard PPM. When $p_{in} = p_{out}$ this is the ER model, so this makes an appropriate null. We fix $p_{in} = 0.8$ and for various values  $p_{out} \leq p_{in}$ construct $N_G = 100$ networks with $3$ equally sized groups of $20$ nodes using the PPM. On each of these networks we determined pseudo p-values according to the procedure in Section \ref{sec:algo}, using the ER model as the null and performing $N_P = 100$ simulations per network. The data points in Figure \ref{fig:er}(a) show the results. The PPM groups can be detected only below the Kesten-Stignum bound, which measures the signal to noise ratio \citep{decelle2011asymptotic, kesten1966limit},
\begin{equation}
    (p_{in} - p_{out})^2 =\frac{Cp_{in} + C(C-1)p_{out}}{N}
\end{equation}
Figure \ref{fig:er}(a) shows the value of $p_{out}$ at the boundary is exactly at the value where the p-value increases sharply and the planted structure is no longer statistically significant.

Figure \ref{fig:er}(b) shows a `problem case' identified by \citep{peixoto2023descriptive} where traditional modularity maximisation over-fits. The network is an ER random graph with $N=30$ nodes and connection probability $p=0.1$ where we have added extra edges to the first $N_{clique}$ nodes to make a clique. Call this the `planted clique model'. We again use the ER null, $N_G = 100$, $N_P=100$ and this time 
$$B = \begin{pmatrix}
1 & 0 \\
0 & 0
\end{pmatrix}$$
to detect one community among unassigned nodes. Before adding the clique the mean degree is $\bar{k} \sim 3$ and the standard deviation is $\sigma \sim 0.4$. Figure \ref{fig:er}(b) shows that when the degree of the nodes in the clique, $N_{clique}-1$, is two sigma or more greater than $\bar{k}$ we can identify significant community structure. The inset shows that the clique nodes are the ones grouped in a community by maximising $Z$.

\begin{figure}[H]
    \centering
    \includegraphics[width=\textwidth]{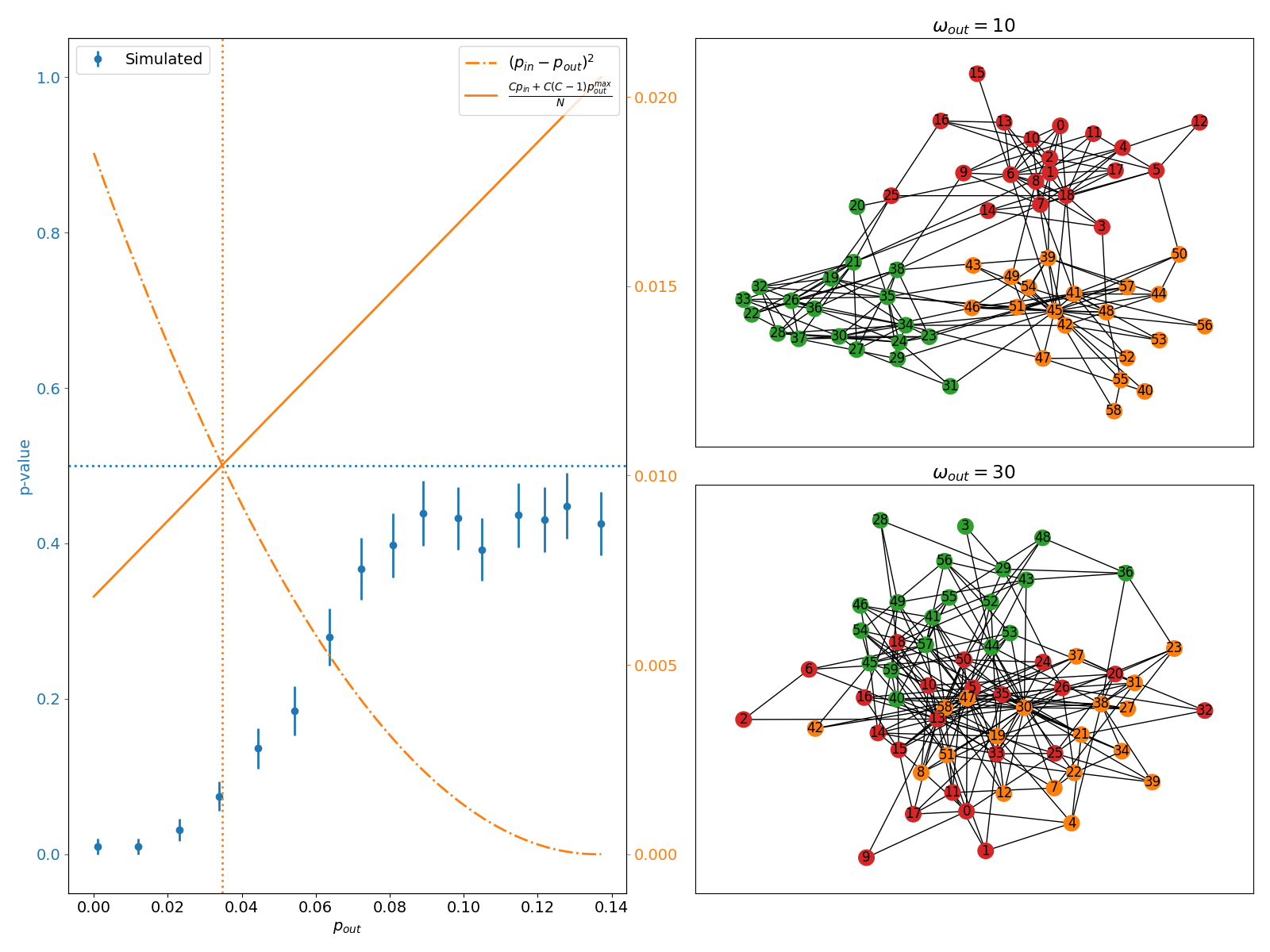}
    \caption{Left shows p-values for detection of a 3 community dc-PPM against the canonical configuration model. The vertical dashed line shows where the Kesten-Stigum bound is broken. The horizontal dashed line shows a p-value of 0.5. On the right are community partitions found by maximising $Z$, when the intercommunity edges are sparse, $\omega_{out} = 10$, and when they are denser, $\omega_{out} = 30$. Communities are planted so that indices $0$ to $19$ are in community 1, $20$ to $39$ in 2 and $40$ to $59$ in 3.}
    \label{fig:cfg}
\end{figure}
Figure \ref{fig:cfg} shows the same experiment using the dc-PPM with  $\omega_{in} = 150$ and $\gamma=3$. We use the canonical configuration model as the null and $N_G = 100$, $N_P=100$ as before. This means we are looking for clustering that is not explained by the degree sequence and find significant community structure cannot be detected above the Kesten-Stigum bound. This does not mean the planted partition could not be recovered by some method. Although the transitions are sharp there is still a weight of evidence for structure in the observed graph, with a p-value $<0.5$, so a partition at least \emph{correlated} with the planted one could be recovered \citep{decelle2011asymptotic, gulikers2018impossibility}. What a non-significant result (say $p>0.01$) means is that there is nothing unusual about this graph in the canonical configuration model. Many networks with the same degree sequence will have a similar community structure. See \citep{arthur2025detectability} for other examples where a hypothesis testing and information theoretic approach give different views on the same network.

\subsection{Random Dot Product Graph Null}\label{sec:rdpgexample}
\begin{figure}[H]
    \centering
    \includegraphics[width=\textwidth]{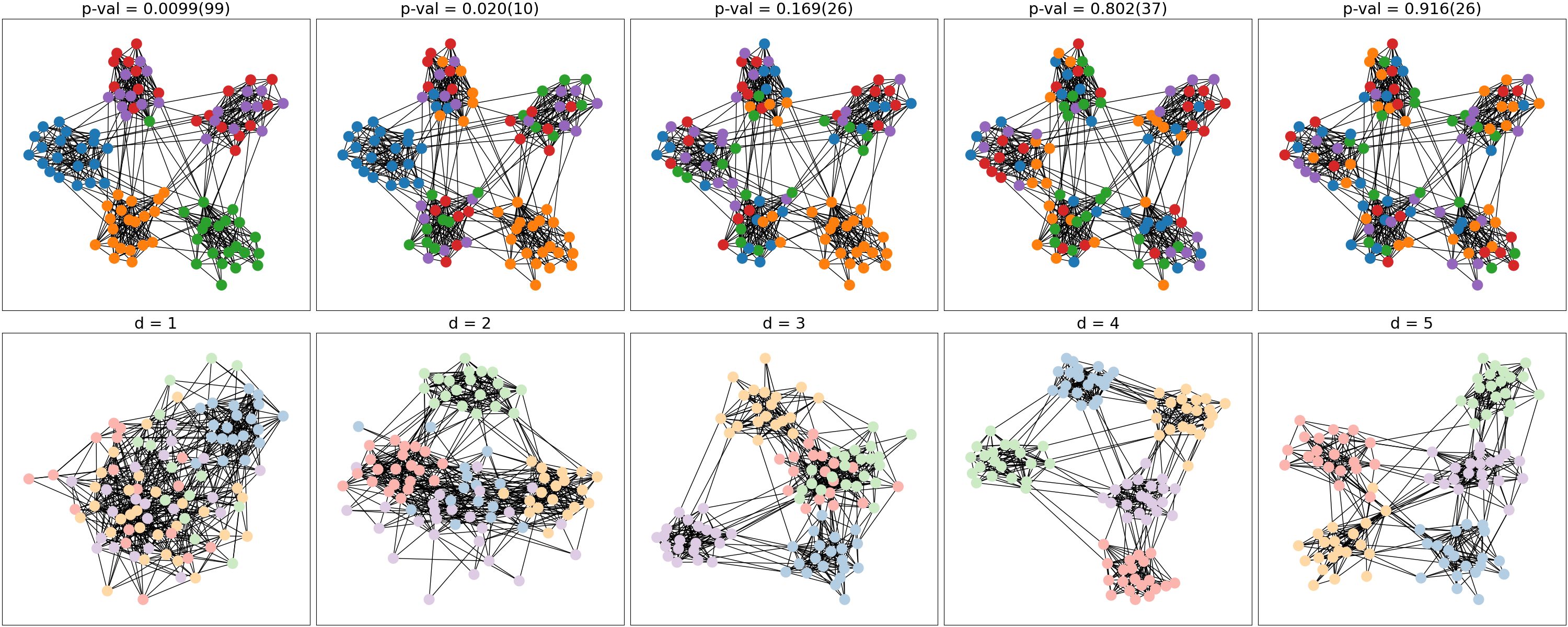}
    \caption{Top row shows a network generated with the dc-PPM, with communities detected using the RDPG null using the top $d = 1, 2, \ldots 5$ eigenvectors as constraints. Bottom row shows an example network generated by sampling edges from the corresponding null and colouring based on communities detected with the canonical configuration model.}
    \label{fig:3}
\end{figure}
To test the RDPG null we generate a network with 100 nodes and 5 equally sized communities using the dc-PPM with $\omega_{in} = 500$, $\omega_{out} = 5$ and $\theta$ chosen by sampling from a Pareto distribution with $\gamma = 3$. We use the RDPG null, Equation \ref{eqn:rdpgnull}, with $v_{m}$ equal to the largest $d$ eigenvectors to detect communities. The p-values labelling the top row of Figure \ref{fig:3} show that for $d = 1, 2, 3$ there is still evidence for `unusual' community structure in the observed network compared to the null. For $d = 4, 5$ there is no strong evidence for more community structure in the observed network than in a random sample from the null. This implies that 4 eigenvectors are enough to resolve the clusters in this network.

The bottom row of Figure \ref{fig:3} shows an example network generated from each null. As the force directed layout and the colours imply, there is weak evidence for 5 clusters for the $d=1,2,3$ networks while the community structure for $d=4,5$ is `obvious'. The high p-values for $d\geq4$, which are averages over $N_G=100$ different networks, mean that the observed network doesn't have more community structure than networks like the ones in the bottom row.

\subsection{Gravity Null}
\begin{figure}[H]
    \centering
    \includegraphics[width=\textwidth]{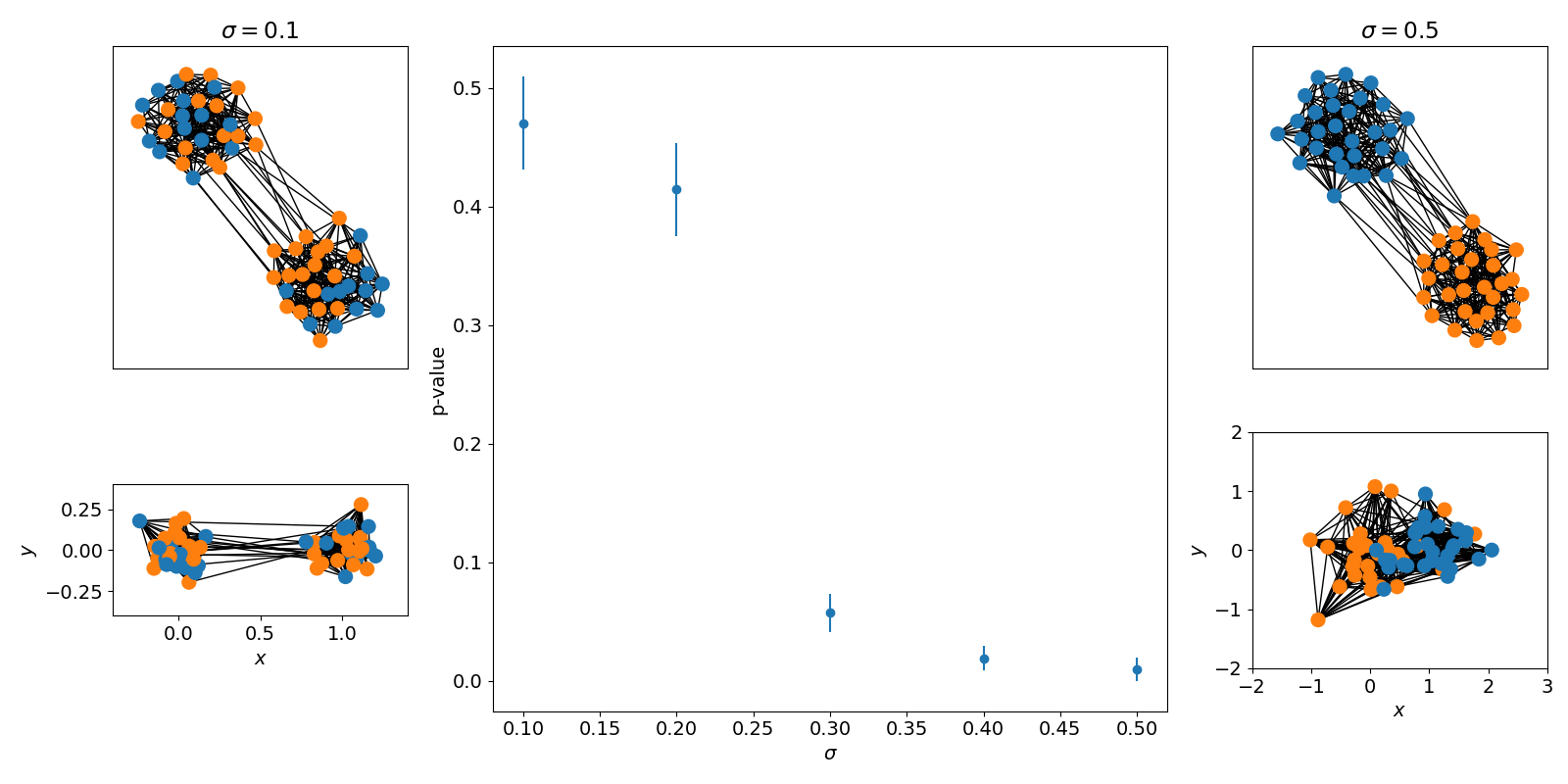}
    \caption{In the left and right panels the top shows the network's topological layout and the bottom shows the spatial layout. The central scatter plot shows p-values versus $\sigma$. When $\sigma$ is small the spatial positions imply clustering will happen in the gravity model. Thus there is nothing `unusual' about the planted communities, clustering is expected when nodes are close. When $\sigma$ is large the gravity model implies less clustering within the groups than we have planted, so significant community structure can be detected.}
    \label{fig:grav}
\end{figure}
We use the PPM with $p_{in} = 0.5$ and $p_{out} = 0.025$ to generate $N_G = 100$ networks with two blocks and $30$ nodes per block. We then locate each of the blocks by assigning a random Gaussian position, $(x,y) \sim ( {\cal N}( \mu_{x}, \sigma), {\cal N}( 0, \sigma) )$, where $\mu_x = 0$ for the first block, $\mu_x = 1$ for the second. When $\sigma$ is small, the blocks are spatially well separated and the null predicts many edges between nearby nodes and few between distant nodes. Community detection with the gravity null gives a high p-value, so we fail to find significant clustering beyond what is expected given the distance between nodes. On the other hand, when $\sigma$ is large, the two planted communities overlap spatially and the spatial position alone does not imply strong clustering into two groups. Thus the excess of connections between the groups is more than the gravity model would suggest and the planted communities are significant.

\section{Comparison with SBM Inference}\label{sec:SBM}

An increasingly popular approach to network community detection is Bayesian inference \citep{peixoto2014hierarchical, peixoto2019bayesian, peel2022statistical}. The basic idea is to maximise the posterior $P(c|A)$, which expresses the probability that the partition $c$ generated the data $A$ according to an underlying generative model, usually some variant of the SBM. Finding the partition which maximises the posterior probability of the data is equivalent to minimising the description length, the number of bits required to specify the data given the model \citep{rissanen1983universal}. The `best' partition according to this approach is the  \emph{maximum a posteriori} (MAP) estimator $\hat{c} = \argmax_c P(c|A)$. This partition also minimises the description length, hence this approach is often referred to as the minimum description length (MDL) principle. For more details see the cited references.

This approach, broadly speaking, answers the question \emph{If my data was generated by an SBM, what were its parameters?} The parameters usually of most interest are the block membership labels, which partition the nodes.  The SBM and its variants are quite flexible and many networks that interest network scientists could be plausibly generated by one. The MDL principle therefore represents a reasonable approach to unsupervised clustering on networks.

The approach described in Section \ref{sec:algo} is Frequentist rather than Bayesian inference. Instead of estimating distributions of model parameters, as with Bayesian methods, we consider our observed network as one possibility chosen from a large set of potential networks generated by the null. If our null model has very low probability to generate the observed network we can rule out the possibility that the observed network came from the null. Rather than an unsupervised learning task, or model fitting, this approach is most suitable for answering specific questions about the data, with assumptions made explicit through the choice of null. 

Another contrast between our procedure and the Bayesian approach is that the Bayesian method uses hard constraints \citep{peixoto2017nonparametric}. This means when computing $P(c|A)$ by
\begin{equation}
    P(c|A) = \frac{P(c, A)}{P(A)}
\end{equation}
$P(c, A)$ is marginalized using
\begin{equation}
    P(c,A) = \sum_\pi P(c,A,\pi) = P(c,A,\hat{\pi})
\end{equation}
where $\pi$ represent model parameters. The formula above means that instead of summing over all possible model parameters weighted by probability, only one term is used to evaluate the sum, the term using the parameters $\hat{\pi}$ that correspond to the values computed on the observed network i.e. sums are over the \emph{microcanonical} ensemble.

Ensemble equivalence means, in the limit of large $N$, averages computed on the microcanonical and canonical ensembles agree. This does \emph{not} hold in general. In the constraint based, entropy maximisation formulation, canonical and microcanonical ensembles are not equivalent when the number of constraints is extensive (proportional to $N$) \citep{squartini2015breaking}. \cite{garlaschelli2017ensemble} show that for the standard SBM we do have ensemble equivalence, but not for the dc-SBM, which is typically used in Bayesian community detection. The canonical approach is suitable when there is the possibility of noise in the data e.g. missing links. Since this is the case for most network data of interest, the canonical ensemble is the appropriate one. We also note recent work \citep{giuffrida2025description} showing that when an extensive number of constraints are enforced, different choices of priors can lead to large differences in description length, of order $\Theta(N)$.

The remainder of this section will compare the Bayesian and frequentist approaches on some well-known networks.

\subsection{Karate Club}
\begin{figure}[H]
    \centering
    \includegraphics[width=\textwidth]{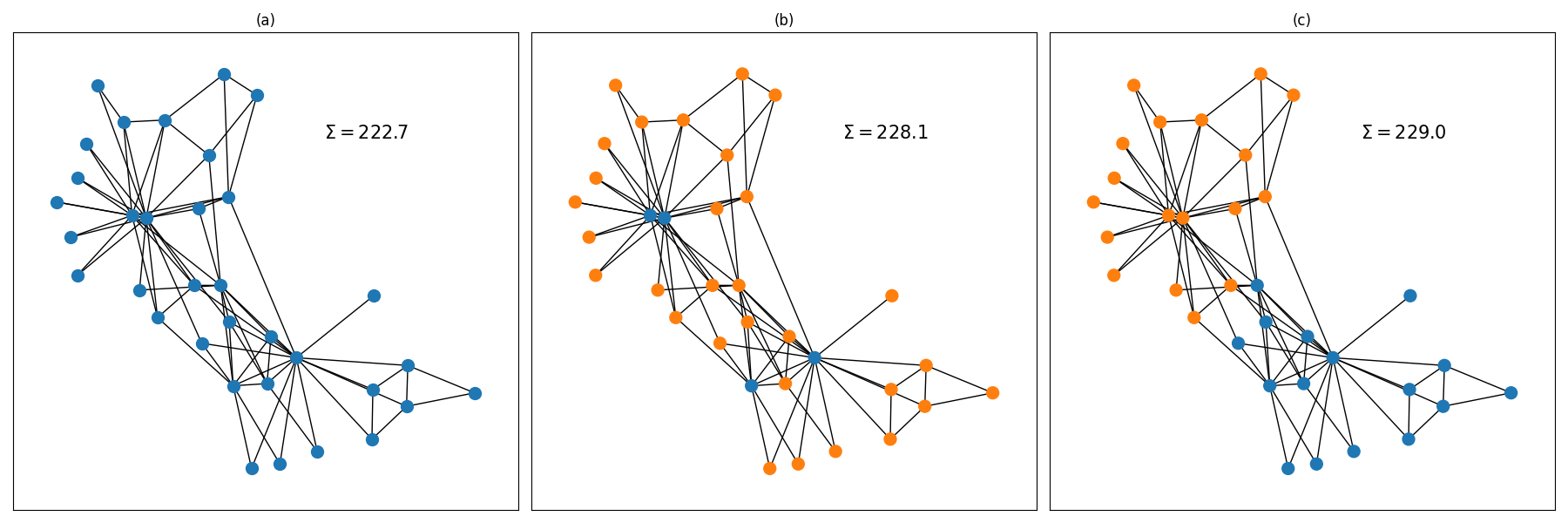}
    \caption{Partitions of the karate club network found with dc-SBM inference. Left is the MAP estimator, indicating no communities. Two other high probability partitions are shown with the description length $\Sigma$ indicated (measured in nats, $\log_e$).}
    \label{fig:sbmkarate}
\end{figure}
Figure \ref{fig:sbmkarate} shows the well-known karate club graph \citep{zachary1977information}, a network of social interactions between members of a karate club just prior to a split caused by a conflict between the club's instructor and president. The analysis of this network from  \cite{peixoto2019bayesian} is repeated using the graph-tool library \citep{peixoto2014hierarchical}. The MAP estimate, using the dc-SBM as the underlying model, is that no community structure is supported, the karate club is consistent with a random graph with the specified degree sequence. If we soften the MDL principle (the small description length principle perhaps) and force $K=2$ communities, we find that a `leader-follower' partition and a two community split are high probability partitions. Given these three rather different possibilities, it is not clear what one should conclude about the community structure of the karate club. Following the MDL principle strictly, there is no community structure, contradicting what is known about the social interactions encoded by the network. Being less strict and looking at other high probability partitions, we have two completely different labellings of the network with similar description lengths.

\begin{figure}[H]
    \centering
    \includegraphics[width=\textwidth]{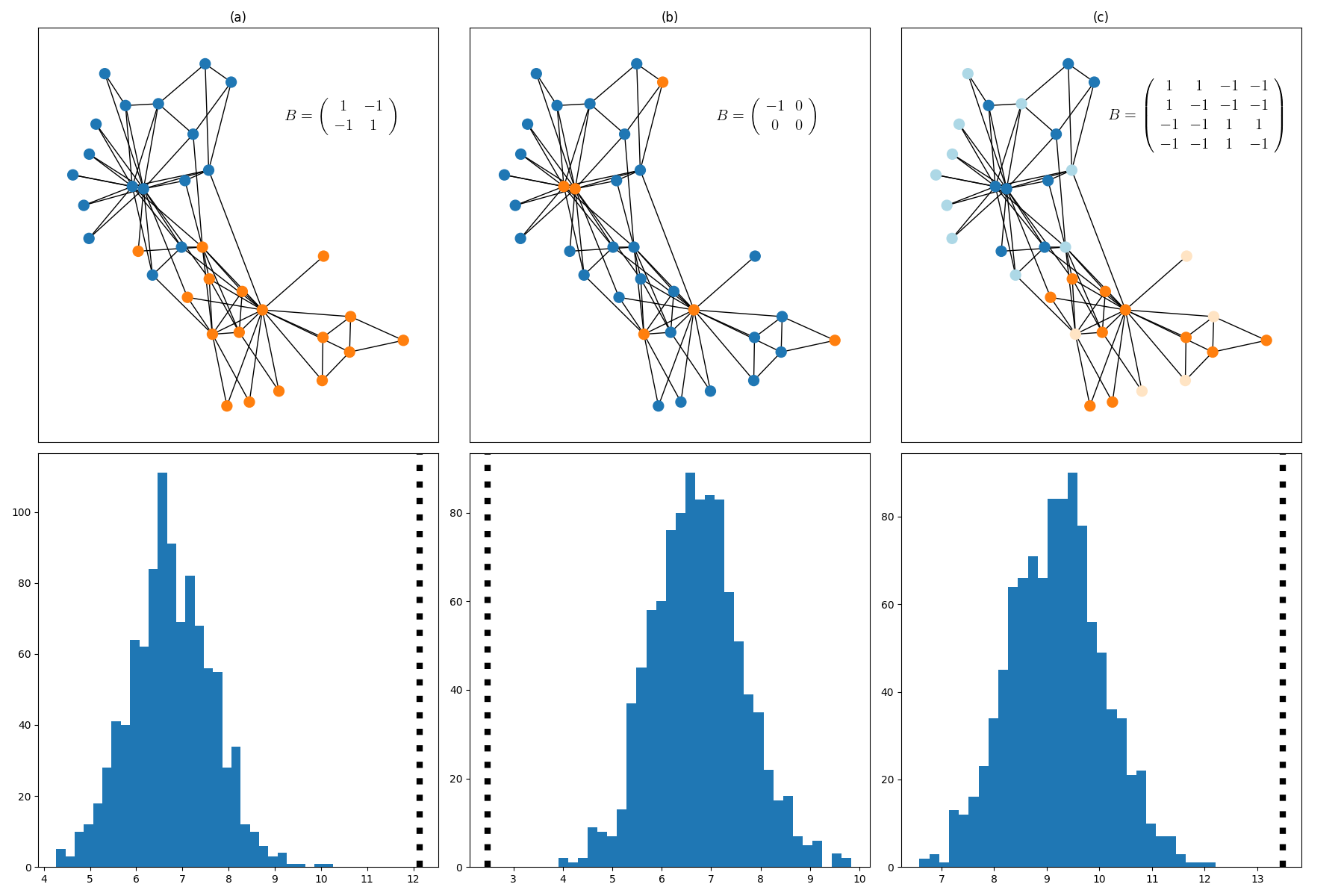}
    \caption{(a) Partitions of the karate club network found from optimising $Z$ against the canonical configuration model with different $B$. Left shows a two community partition. (b) a `repulsive' partition. (c) a double core-periphery. Bottom: shows the distribution of $Z$ values from the null, with the dashed line corresponding to the value $\hat{Z}$ computed on the observed network.}
    \label{fig:qkarate}
\end{figure}
Our analysis follows the procedure in Section \ref{sec:algo}. We want to know if the social interactions encoded by the network reflect the conflict between the instructor and president, or if they are simply due to the fact that these individuals have a lot of friends. We translate this to the question: \emph{does the observed network have stronger community structure than any other network with the same degree sequence?} Here `strength' is measured by $Z$ and the null ensemble enforces the degree constraint \emph{on average}. 

Doing the optimisation and simulation the answer is Yes. Figure \ref{fig:qkarate}(a) shows that in 1000 samples from the null we never observe a network with such strong community structure. This holds even if we allow an `unaligned' partition by appending a row and column of $0$s to $B$ as in Section \ref{sec:unassigned}, the optimal partition does not use the unaligned label, it is better, in terms of optimal $Z$, to put all nodes in one or the other community. 

We can experiment with other structures. A bipartite structure
$$
B = \begin{pmatrix}
-1 & 1\\
1 & -1
    \end{pmatrix}
$$
is not significant, nor is a `repulsive' partition
$$
B = \begin{pmatrix}
-1 & 0\\
0 & 0
    \end{pmatrix}
$$
shown in Figure \ref{fig:qkarate}(b). Interestingly, in this case the observed data has significantly \emph{smaller} $\hat{Z}$ than a random network with this degree sequence. So this network is unusually difficult to split into a group which does not interact and a group which interacts indiscriminately. This strongly implies the `feud' is not restricted to just a few key individuals.

We can still refine this picture. Using a double core-periphery (CP) \citep{kojaku2017finding} block pattern
\begin{equation*}
    B = \begin{pmatrix}
        1 & 1 & -1 &-1\\
        1 & -1 & -1 &-1\\
        -1 & -1 & 1 &1\\
        -1 & -1 & 1 &-1\\
    \end{pmatrix}
\end{equation*}
In 1000 samples from the null we find no networks with equal or stronger double CP structure. The partition is shown in Figure \ref{fig:qkarate}(c) and suggests individuals who may only be peripherally involved in the conflict.

\subsection{Political Blogs}

\begin{figure}[H]
    \centering
    \includegraphics[width=\textwidth]{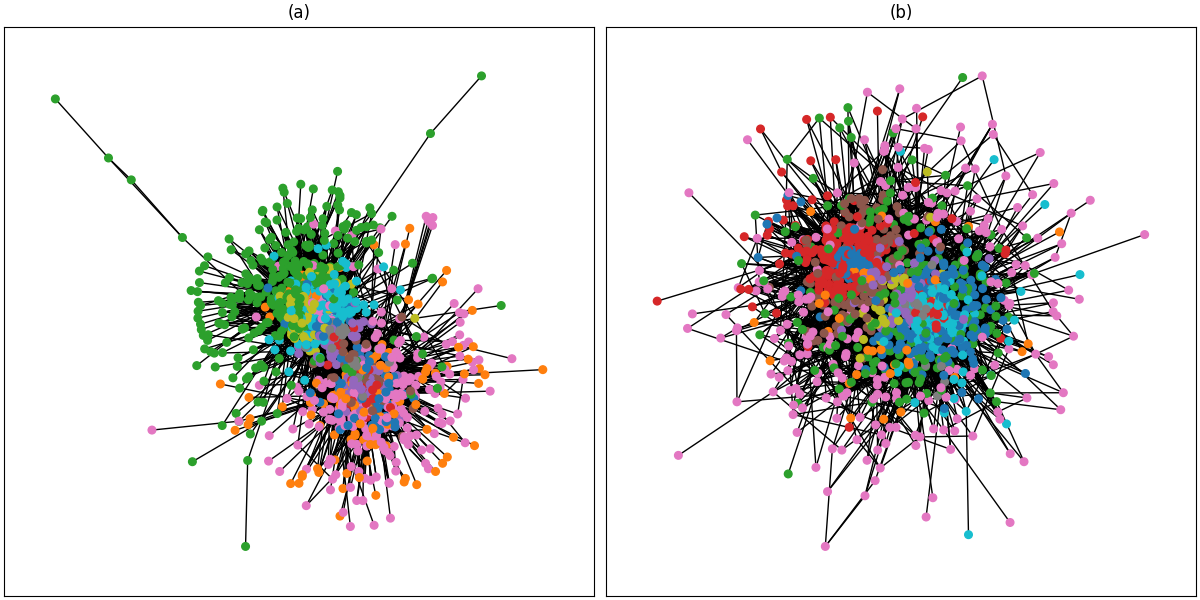}
    \caption{(a) SBM based minimum description length partition of the political blogs network. There are 19 distinct groups. (b) MDL partition of a random network from the 2 eigenvector RDPG null model. This has 15 groups. }
    \label{fig:polblogssbm}
\end{figure}
Figure \ref{fig:polblogssbm}(a) shows a network of links between political blogs (treated as undirected) \citep{adamic2005political}. The force directed layout agrees with intuition that there should be two main clusters (left and right wing blogs), although they appear to be strongly overlapping. The MAP estimate suggests 19 communities, which are hard to interpret without any additional metadata.

\begin{figure}[H]
    \centering
    \includegraphics[width=\textwidth]{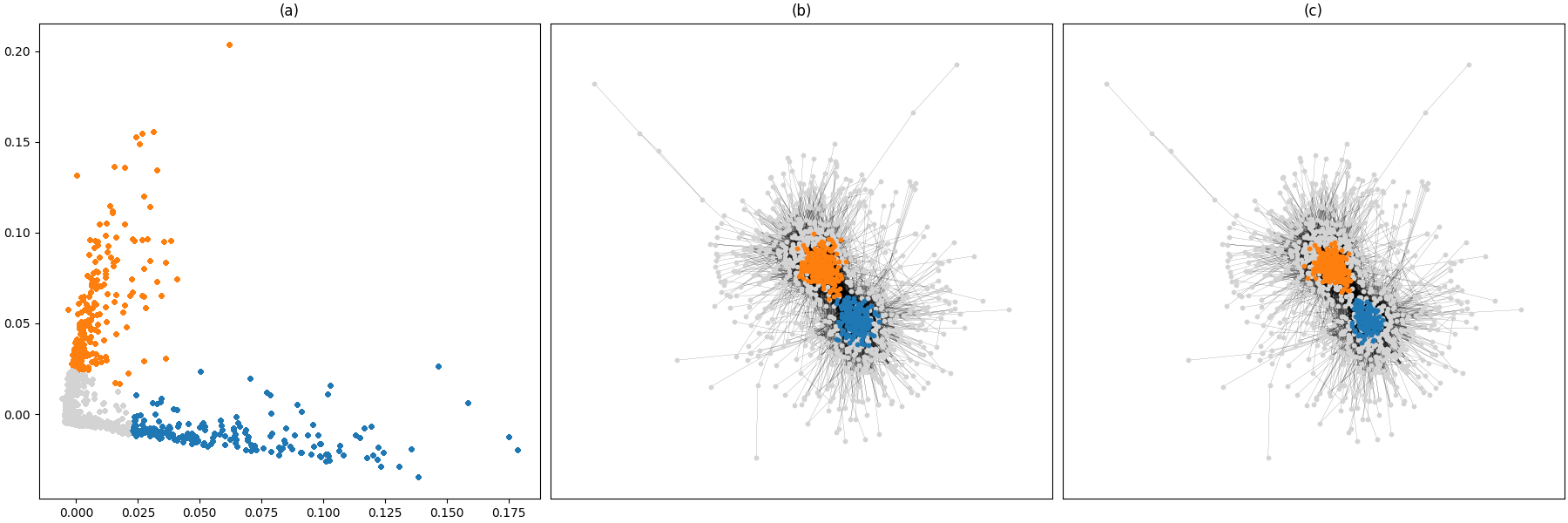}
    \caption{(a) Vector representation of the poltical blogs data using the two largest eigenvectors. Colours derived from vector based clustering algorithm. (b) Political blogs network, coloured based on eigenvector clustering. (c) Political blogs network, coloured based on labels maximising $Z$, using the ER null and $B$ equal to a 2 community plus unaligned (grey). }
    \label{fig:polblogs2}
\end{figure}
Figure \ref{fig:polblogs2}(a) shows a vector representation of each node using the first two eigenvectors of the adjacency matrix. We can cluster this vector data using a standard algorithm \citep{von2007tutorial} and use the labels to colour the nodes of the network, as in Figure \ref{fig:polblogs2}(b), suggesting the left/right split and a large periphery. We can reproduce a similar labelling by maximising $Z$ under the ER null, with the structure matrix 
$$  
B = \begin{pmatrix}
1 & -1 & 0\\
-1 & 1 & 0\\
0 & 0 & 0
    \end{pmatrix}
$$
Shown in Figure \ref{fig:polblogs2}(c). 

We will now construct a RDPG model which fixes the first two eigenvectors, i.e. $A$ and $P$ have the same rank-2 approximation. We can use this to generate networks which have the same rank-2 approximation as the political blogs data. Figure \ref{fig:polblogssbm}(b) shows such a network along with its MAP partition, obtained using graph-tool. This suggests the generated network has a partition of 15 groups. Given the network is randomly generated and contains no more information than Figure \ref{fig:polblogs2}(a), the MAP partition is most likely spurious, or at least very difficult to interpret.

\begin{figure}[H]
    \centering
    \includegraphics[width=\textwidth]{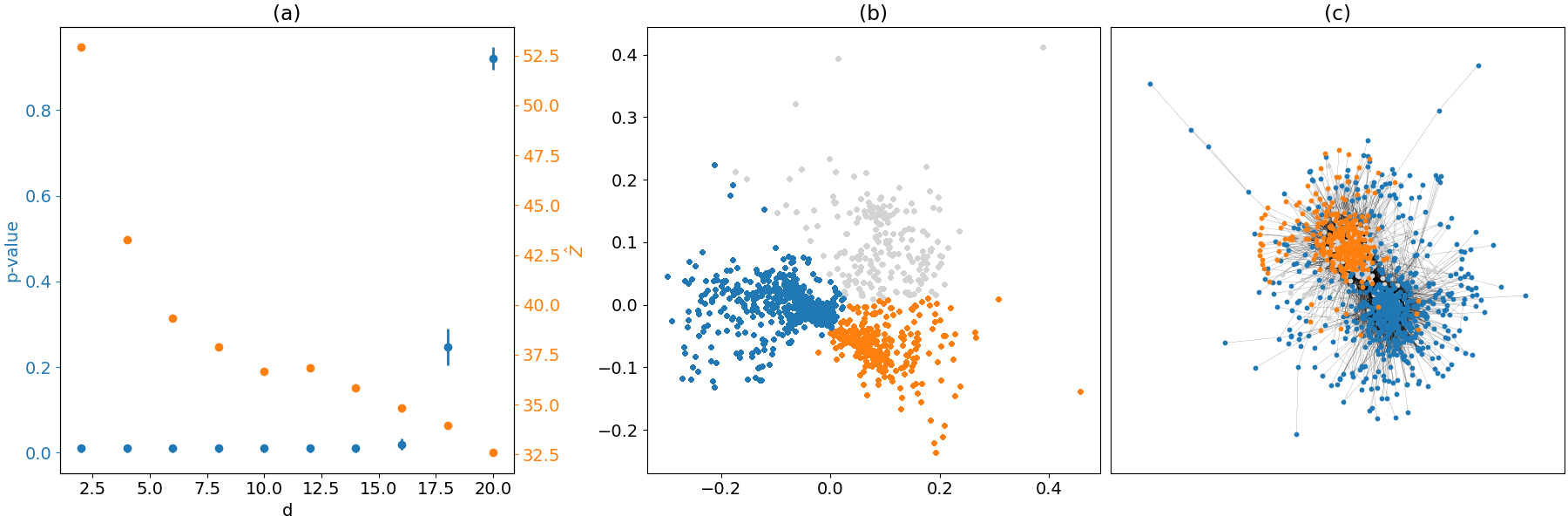}
    \caption{(a) p-value of the 2 community plus unaligned partition using an RDPG null with $d$ eigenvectors. Orange shows the value of $\hat{Z}$, blue the p-values. (b) Clustering nodes based on the highest 20 eigenvectors. Data is projected to 2d using the Isomap algorithm \citep{tenenbaum2000global} for the purposes of visualisation, but is clustered in 20 dimensions. (c) Political blogs network, coloured based on labels determined in (b). }
    \label{fig:polblogs20}
\end{figure}
As in Section \ref{sec:rdpgexample}, we can explore how many vectors are required to `explain' all of the community structure in the political blog network. We construct $P_{ij}$ as in Section \ref{sec:rdpgexample} using $d = 2 \ldots 20$ eigenvectors and measure the significance of partitions found using the 2 community plus unaligned structure matrix $B$. Figure \ref{fig:polblogs20}(a) shows that $\sim 16$ eigenvectors is enough to `explain' all the structure i.e. this is the point at which networks from the null become statistically indistinguishable, as measured by $\hat{Z}$, from the observed network. Figure \ref{fig:polblogs20}(b) uses the top 20 eigenvectors to cluster the data and Figure \ref{fig:polblogs20}(c) shows these cluster labels applied to the network. This gives two large communities and a few unaligned nodes. This is much more in line with intuition, and with the metadata on affiliation presented in \cite{adamic2005political}, than the MDL partition into 19 groups. It can also answer the most likely questions we have about the data i.e. is this blog left-wing, right-wing or neither?

\section{Discussion}\label{sec:conclusion}

Most discussions of community detection frame it as \emph{exploratory data analysis}. We have a network and are completely in the dark as to what kinds of groups or structures it might contain. We contend that this is not the only, or even the most common, use case for community detection. We often have metadata about the nodes e.g. a social network of students where we know the courses the students are taking. We also often have specific questions about the network e.g. do friendship groups reflect political party affiliation? One approach is to incorporate this metadata into the community detection algorithm \citep{peel2017ground, emmons2019map}. In this work rather than guiding the exploration with metadata, we suggest a move from exploratory to \emph{confirmatory data analysis} and formal hypothesis testing. 

For exploratory analysis one should define what is meant by a `community', which is what different algorithms do, explicitly or implicitly, e.g. communities are connected by bridges \citep{newman2004finding}, communities are places where random walkers get stuck \citep{rosvall2008maps}, communities allow us to effectively compress the network \citep{peixoto2019bayesian} etc. In this paper a community is defined by maximising $Z$, so for us a community partition maximises the z-score between observed edge counts and expected edge counts under the null. Inequivalent notions of community might not necessarily reproduce the same partitions. This does not mean one algorithm is always better or results should be averaged to form a consensus. It means different labelling algorithms are looking for different things and the analyst must be precise about what they mean by a `community'. On the other hand this does not mean that all algorithms are equally valid, and some approaches, in particular standard modularity maximisation \citep{peixoto2023descriptive}, have serious problems and should be avoided when good alternatives are available.

However, finding labels using some algorithm is not the final step. In this paper we have discussed performing confirmatory data analysis by comparing the observed network to other networks in a null model. We particularly emphasise using null models which are mathematically sound i.e. constructed from correctly normalised probabilities that sample the canonical ensemble. Nulls should also be sufficiently powerful. In medical testing it is easier to demonstrate a significant effect by comparing a new medicine to a placebo than by comparing it to the current best treatment. Similarly for community detection, some structure might be significant under, say, an ER null, but not under a configuration null. Or it might be significant under both but not under an RDPG null fixing the first few eigenvectors. 

The examples shown in Section \ref{sec:artificial} and Section \ref{sec:SBM} demonstrate this and counter the contention of \citep{peixoto2023descriptive} that hypothesis testing approaches to community detection always demonstrate significance and offer little insight. As also shown in Sections \ref{sec:artificial} and \ref{sec:SBM} the partition found depends on the null. Another excellent example (albeit using modularity) is given by \citep{expert2011uncovering}, where using the configuration model communities correspond to geographic regions, and using a spatial null communities correspond to linguistic groups. The `correct' partition depends on the question you are asking. Such spatial null modelling fits well in our framework \citep{di2022gravity} and will be explored more in future work.

Exploratory and confirmatory data analysis are not mutually exclusive, they are complimentary. Our approach requires formulating precise research questions. Formally this means picking $P, B$ and $K$. Specific research hypotheses, metadata or exploratory data analysis, perhaps involving `unsupervised' community detection, are all reasonable ways to suggest choices of $P$, $B$ and $K$. Exploratory data analysis is important and can give useful insight into network data. But, as in most fields of science, strong conclusions require carefully formulated hypotheses and quantified uncertainty. This applies just as much to network community detection as it does to the measurement of any other quantity. Rather than relying on the outputs of some black-box algorithm, or worse, averaging together the outputs of many black-boxes, practitioners should be encouraged to make testable and concrete hypotheses about the structure of their networks. Methods like those presented here can then test such hypotheses and generally move us towards a more rigorous approach to community detection. 

\bibliographystyle{plainnat}
\bibliography{references}

\end{document}